\documentclass[12pt,a4paper]{article}
\usepackage{lineno}
\usepackage{graphics}
\usepackage{graphicx}
\usepackage{subfigure}
\usepackage{amsmath}
\usepackage{amssymb}
\usepackage{authblk}
\title{Design and realization of a facility for the characterization of Silicon Avalanche PhotoDiodes}

\author[1]{Andrea Celentano\thanks{Corresponding author: andrea.celentano@ge.infn.it}}
\author[2]{Luca Colaneri}
\author[1]{Raffaella De Vita}
\author[1]{Stuart Fegan}
\author[1]{Giuseppe Min\`i}
\author[2]{Gianni Nobili}
\author[1]{Giacomo Ottonello}
\author[1]{Franco Parodi}
\author[2]{Alessandro Rizzo}
\author[1]{Irene Zonta}

\affil[1]{INFN, Sezione di Genova\\
Via Dodecaneso 33, Genova, Italy}
\affil[2]{INFN, Sezione di Roma Tor Vergata\\
Via della Ricerca Scientifica 1, Roma, Italy}

\begin{document}
\date{Published in: JINST 9, T09002, 2014}
\maketitle
\begin{abstract}
We present the design, construction, and performance of a facility for the characterization of Silicon Avalanche Photodiodes in the operating temperature range between -2 $^{\circ}$C and 25 $^{\circ}$C. The system can simultaneously measure up to 24 photo-detectors, in a completely automatic way, within one day of operations. The measured data for each sensor are: the internal gain as a function of the bias voltage and temperature, the gain variation with respect to the bias voltage, and the dark current as a function of the gain. 
The systematic uncertainties have been evaluated during the commissioning of the system to be of the order of $1\%$. 

This paper describes in detail the facility design and layout, and the procedure employed to characterize the sensors. The results obtained from the measurement of the 380 Avalanche Photodiodes of the CLAS12-Forward Tagger calorimeter detector are then reported, as the first example of the massive usage of the facility.
\end{abstract}

%\linenumbers

\section{Introduction}\label{sec:introduction}

Silicon Avalanche PhotoDiodes (APDs) have become in recent years a convenient and common photo-detector choice because of the compact size, good radiation hardness, and insensitivity to intense magnetic fields.
Avalanche PhotoDiodes replaced traditional photo-multipliers tubes in many particle physics detectors. APDs have been successfully used in the CMS detector at the CERN LHC, employing more than 150k APDs in the electromagnetic calorimeter \cite{CMS-ECAL}, and in the ALICE Photon Spectrometer (PHOS), equipped with 18k sensors \cite{PHOS-ECAL}. The PANDA detector at FAIR foresees to use more than 15k Large Area APDs for the electromagnetic calorimeter readout, operating them at $-25$ $^{\circ}$C \cite{PANDA-ECAL}. Other significant examples of large-scale detectors employing APDs for light readout include the Forward Tagger Calorimeter in Hall-B at Jefferson Laboratory \cite{FT-CAL} and the HPS-Ecal Calorimeter \cite{HPS-CAL}, both currently under development.

The intrinsic gain and the dark current of  Avalanche PhotoDiodes strongly depends on the bias voltage and on the operating temperature. It is therefore necessary to characterize them before installation in any experimental setup to be able to work at the selected working point.
This is even more critical for detectors employing Avalanche PhotoDiodes for multiple channel readout, as for electromagnetic calorimeters, since differences between working points could induce degradation of the full detector performances. Mapping the gain of each sensor also permits to group together those with close working point and supply them with the same bias voltage, thus reducing the number of independent high voltage channels. 

In this paper we present our realization of a new a facility for the automatic characterization of Avalanche PhotoDiodes in the $-2$ $^{\circ}$C - $25$ $^{\circ}$C temperature range. The facility was first developed to measure the 380 Large Area APDs (model Hamamatsu S8664-1010) employed in the CLAS12-Forward Tagger Calorimeter. This was the first massive usage of the system. The good performances obtained validated the design of the facility, which was used again later to characterize the 516 Large Area APDs employed in the HPS-Ecal detector.

\section{LA-APDs characterization procedure}\label{sec:ApdProcedure}

\subsection{Intrinsic gain measure}

The procedure employed in the facility to measure the APDs internal gain ($G$) at fixed temperature ($T$) as a function of the bias voltage ($V$) is as follows. The dark current $I_{off}$ and the photo-current $I_{on}$  due to a continuous illumination  are measured at different values of the reverse bias voltage with a picoammeter. The gain is derived using the following equation:

\begin{equation}\label{eq:apd1}
G(V)=\frac{I_{on}(V)-I_{off}(V)}{I_{on}(G=1)-I_{off}(G=1)} \; \; \; ,
\end{equation}
where $I_{on}(G=1)$ and $I_{off}(G=1)$ are, respectively, the values of the photo-current and the dark current when the internal gain of the APD is equal to one, i.e. when the bias voltage is sufficiently low so that the avalanche mechanism is not active. Since the procedure employed to calculate the gain involves the normalization to the unitary gain current, the only requirements for the light source are the stability during the measurement time and an intense light emission to provide a significant $I_{on} (G=1)$ current, not lower than $\simeq 10$ nA, to guarantee a good signal over noise ratio in the current measurement.
Being the spectral response of the sensor typically peaked at 420 nm, a standard blue LED can be employed.
\begin{figure}[tpb]
\begin{centering}
\includegraphics[width=0.6\textwidth]{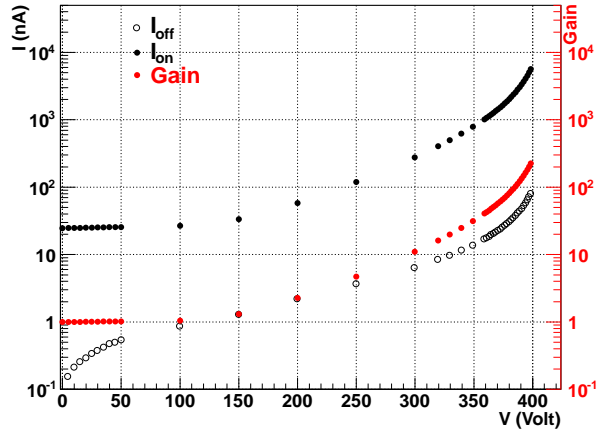}
\caption{\label{fig:apdF1} \small Typical LA-APD photo-current $I_{on}$ (closed black markers) and dark current $I_{off}$ (open black markers) behaviour, as a function of the bias voltage $V$ (these data were obtained measuring a LA-APD, model Hamamatsu S8664-1010). The corresponding internal gain $G$ is also reported (red markers). This measurement was performed at $18.5 \pm 0.1$ $^{\circ}$C, in the voltage range between 0 and 400 V.}
\end{centering}
\end{figure}

\begin{figure}
\begin{center}
\subfigure{\includegraphics[width=0.48\textwidth]{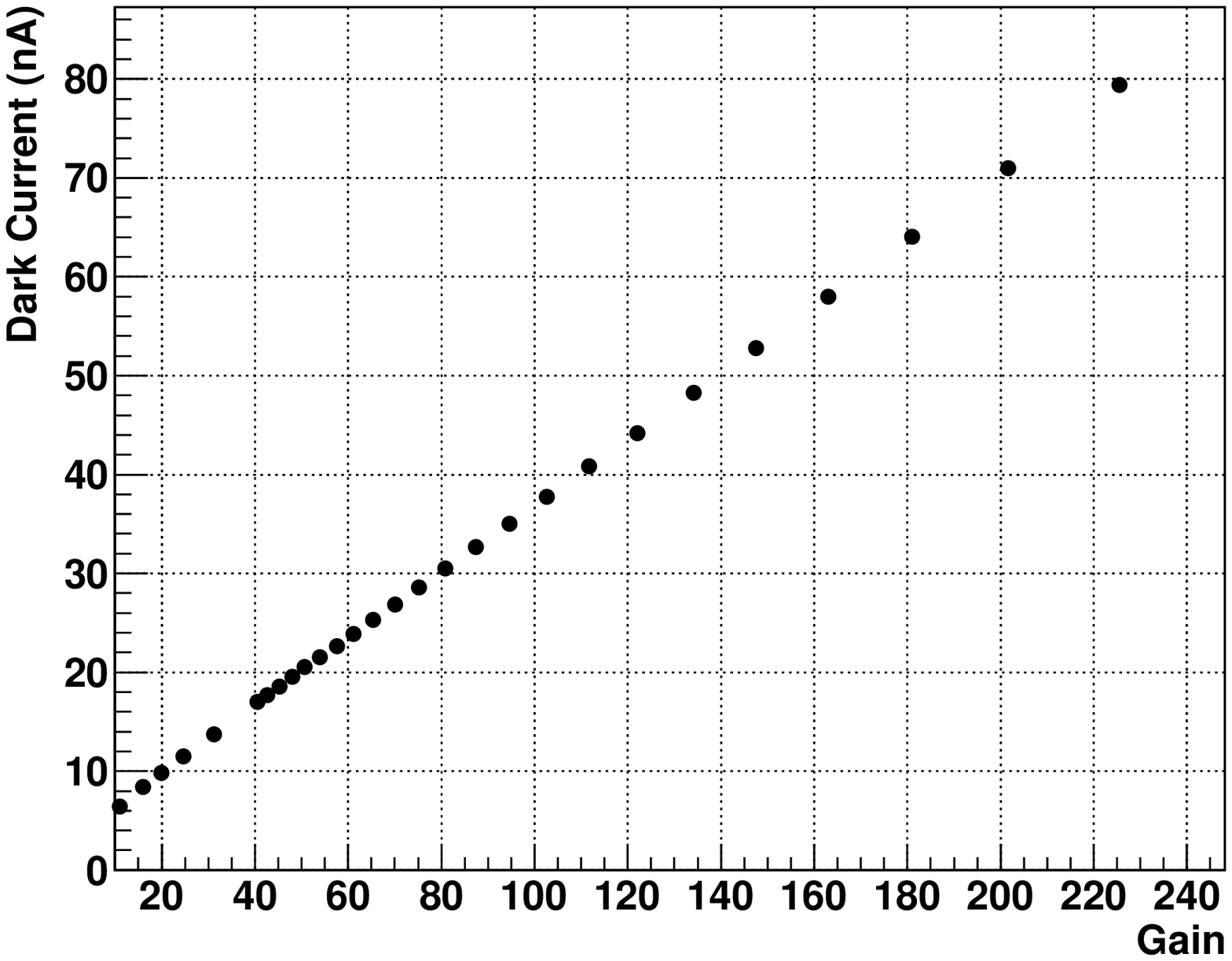}} \quad
\subfigure{\includegraphics[width=0.48\textwidth]{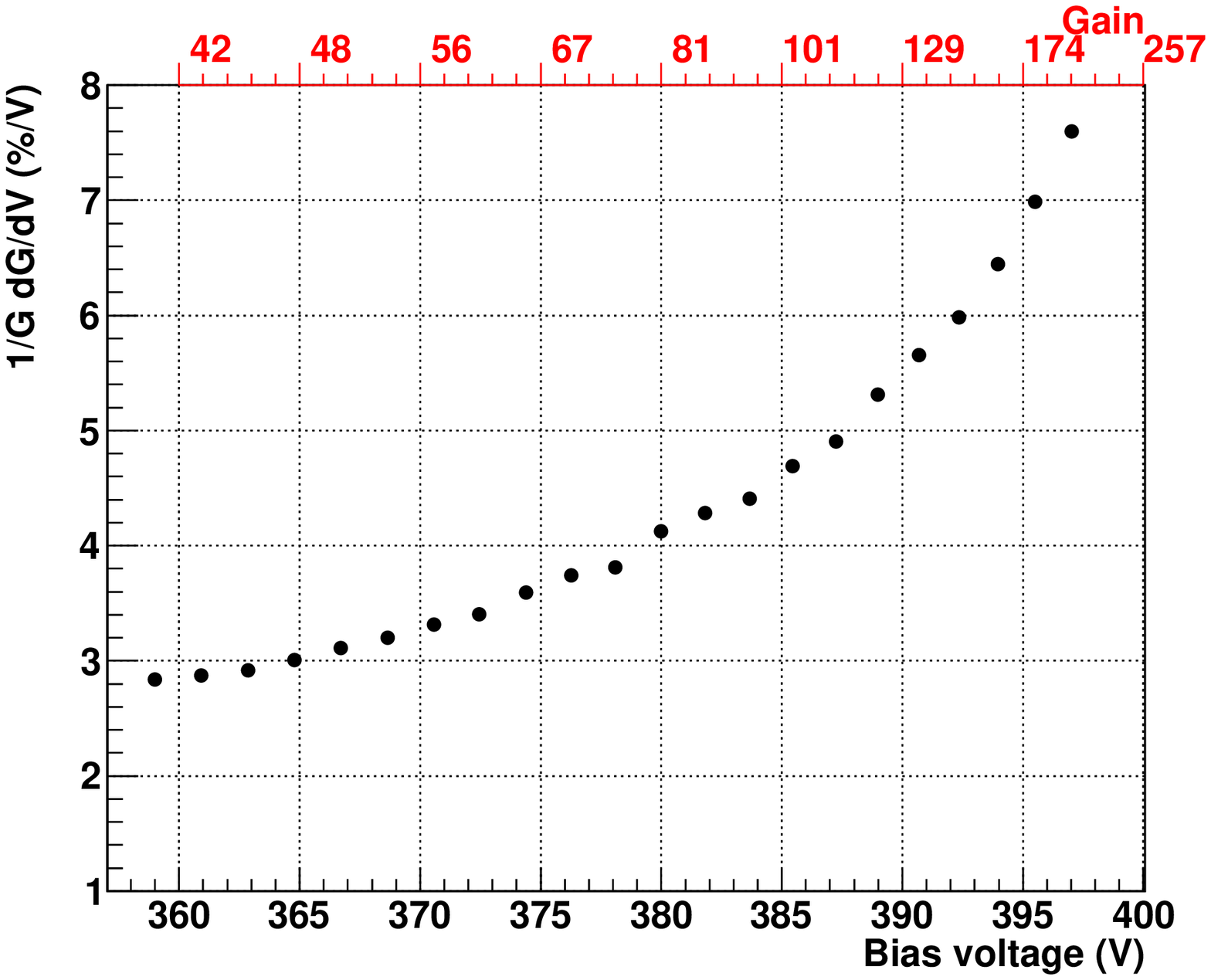}}
\caption{\small Left: LA-APD dark current $I_{off}$ as a function of the LA-APD gain $G$. Right: LA-APD intrinsic gain variation as a function of the bias voltage. The corresponding sensor gain is also reported on the top axis.}
\label{fig:apdF2}
\end{center}
\end{figure}

The typical behavior of an APD photo-current $I_{on}$ and dark current $I_{off}$ is shown in Figure~\ref{fig:apdF1}. After an initial plateau, which corresponds to the unitary gain, $I_{on}$ grows rapidly for increasing bias voltage. $I_{off}$ remains smaller than $I_{on}$ by a factor $\simeq 100 - 1000$. The value of the denominator in equation (\ref{eq:apd1}) thus corresponds to the initial plateau of $I_{on}$. To reduce the measurement time, the voltage scan is performed with non-uniform steps in the above mentioned bias range, with denser measurements in the two regions of interest: the unitary-gain plateau and the neighborhood of the foreseen working point ($G=150$ for the specific example here reported). 
The corresponding internal gain $G$ is also shown in Figure~\ref{fig:apdF1}. The unitary gain plateau ($G=1$) is visible for bias values lower than $\simeq 100$ V, while for bigger values $G$ increases rapidly.
The measured dark current $I_{off}$ is reported in Figure \ref{fig:apdF2}, left panel, as a function of the intrinsic gain $G$. It shows a linear behavior, as expected from the most common interpretation of a LA-APD dark current as a sum of two terms: the first (``surface current'') being independent of $G$, and the second (``bulk current'') being directly proportional to it \cite{PANDA-ECAL}.
%
%A blue LED, with emission peak centered around 420 nm, is employed as light source\footnote{The stability of the LED was checked using a stable PMT and measuring the output current while exposed at the LED light. We did not see any appreciable variation during a measurement period of several hours.}.

The relative APD gain dependence on the bias voltage is evaluated numerically, by computing the corresponding incremental ratio: 
\begin{equation}
\frac{1}{G}\frac{\partial G}{\partial V} \simeq \frac{1}{G(V)} \frac{G(V+\varepsilon)-G(V)}{\varepsilon} \; \; \; ,
\end{equation}
with $\varepsilon=0.1$ V. $G(V+\varepsilon)$ is evaluated from the measured $G(V)$ data points, employing a spline interpolation. A typical result is shown in Figure~\ref{fig:apdF2}, right panel.

%\begin{figure}
%\begin{center}
%\includegraphics[width=0.48\textwidth]{fig3.eps}
%\caption{\small \label{fig:apdF3} LA-APD intrinsic gain variation as a function of the bias voltage. The corresponding intrinsic gain is also reported on the top axis.}
%\end{center}
%\end{figure}

\subsection{Temperature and voltage dependence}

\begin{figure}
\begin{center}
\subfigure{\includegraphics[width=0.48\textwidth]{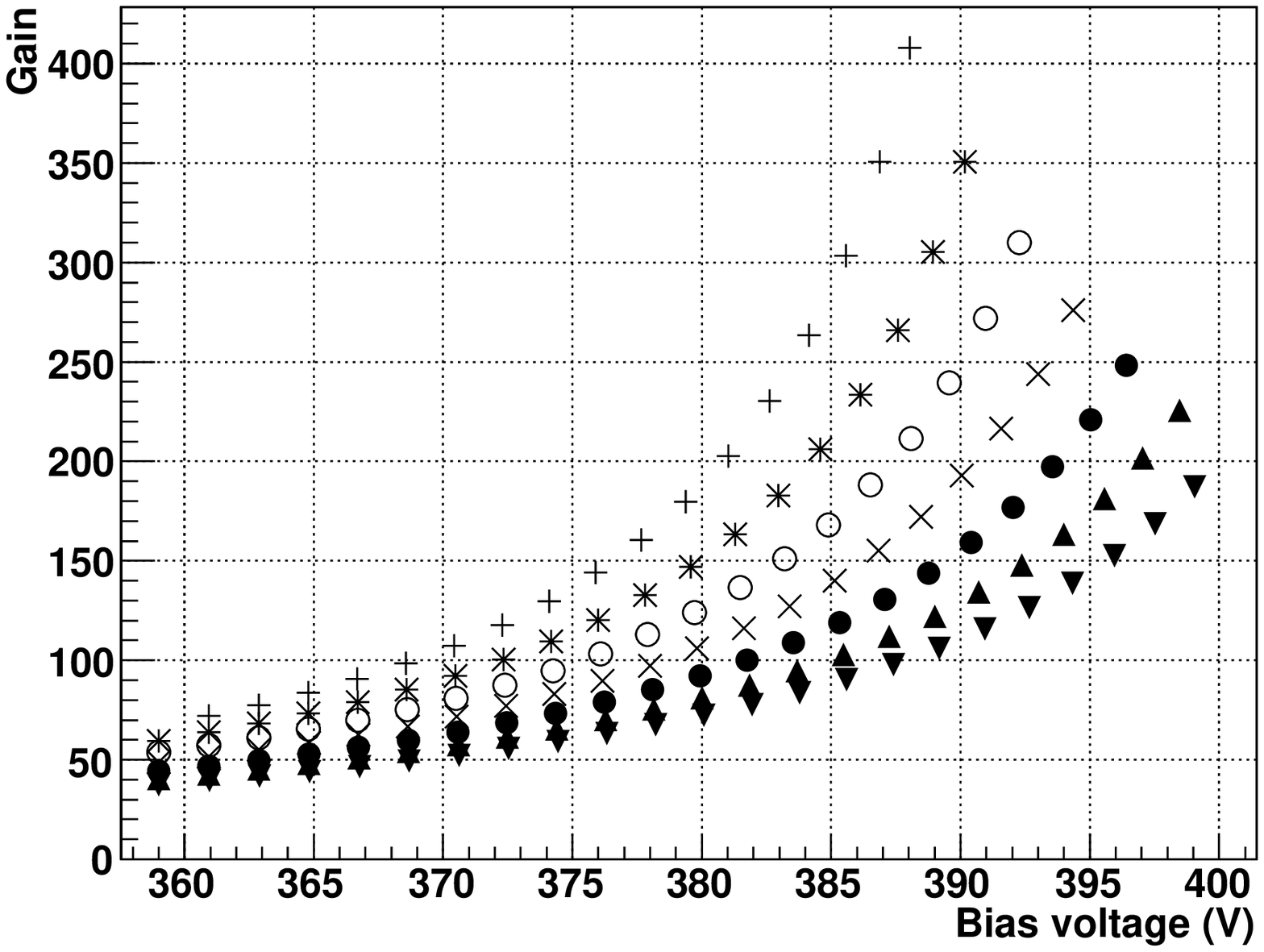}}\quad
\subfigure{\includegraphics[width=0.48\textwidth]{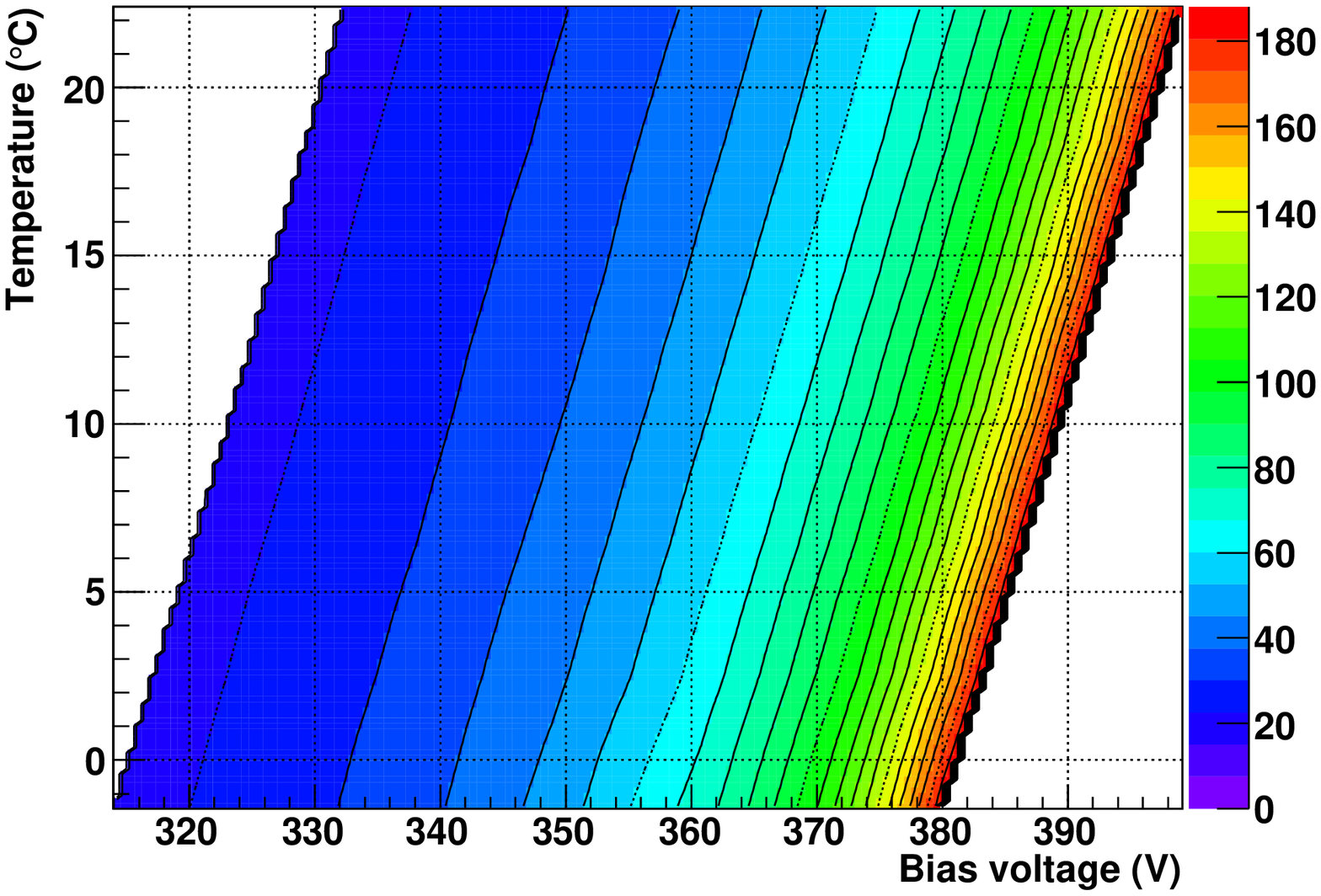}}
\caption{\label{fig:apdF4} \small Left: LA-APD intrinsic gain $G$ for different temperatures. 
$+: -1.44$  $^{\circ}$C, $*: 2.51$ $^{\circ}$C, $\circ: 6.47$ $^{\circ}$C, $\times : 10.44$ ${\circ}$C,
$\bullet: 14.43$  $^{\circ}$C, $\blacktriangle: 18.42$  $^{\circ}$C, $\blacktriangledown: 22.39$  $^{\circ}$C. 
%The open triangle corresponds to the bias declared by the manufacturer for  $G=50$ @ $T=+25^{\circ}$C. 
Right: LA-APD gain (color-scale), as a function of the bias voltage $V$ and operating temperature $T$. The isogain curves are reported in black.}
\end{center}
\end{figure}

The temperature dependence of the APD intrinsic gain and dark current is investigated by comparing measurements performed at different temperatures.
Figure \ref{fig:apdF4}, left panel, shows an example for a sensor characterized in the temperature range between $-1.5$  $^{\circ}$C and $+22.5$ $^{\circ}$C. The same data are reported in Figure \ref{fig:apdF4}, right panel, linearly extrapolated in the whole temperature range. We observed that the isogain curves in the $V-T$ plane are approximately straight lines (in the region where $G>1$, i.e. outside the unitary gain plateua). Therefore, the APD gain can be actually parametrized as a linear combination of these two variables:
\begin{equation}\label{eq:apd2}
G(V,T)=G(\alpha V - \beta T) \equiv G(x) \; \; \; ,
\end{equation}
with $\alpha$ and $\beta$ two proper coefficients to be measured experimentally, and $x=\alpha V - \beta T$.

%\begin{figure}[tpb]
%\begin{centering}
%\includegraphics[width=0.48\textwidth]{fig5.eps}
%\caption{\small \label{fig:apdF5} LA-APD gain (color-scale), as a function of the bias voltage $V$ and operating temperature $T$. The isogain curves are reported in black.}
%\end{centering}
%\end{figure}
We exploited this behavior to derive the bias voltage $V_0$ corresponding to a given gain $G_0$, at temperature $T_0$, as follows.
First, the photo-sensor gain is measured at a few different temperatures $T_i$, and the bias voltage $V_i$ corresponding to a certain fixed gain $G$ is evaluated.
Then, the ratio $k\equiv\alpha/\beta$ is calculated by performing a best-fit with a linear function to the $T_i$ vs $V_i$ data points. The procedure is repeated for different gain values, to check the stability of the parameter $k$. An example is reported in Figure~\ref{fig:apdF6}, left panel. The corresponding value of $k$ is $1.26 \pm 0.01$.
$V_0$ is finally calculated as:
\begin{equation}
V_0 = V_i + k^{-1} \cdot (T_0 - T_i) \; \; \; ,
\end{equation}
where $V_i$ is the bias voltage corresponding to $G_0$ at any of the temperatures $T_i$.

The intrinsic gain relative variation with respect to the operating temperature follows from Eq.~\ref{eq:apd2}:
\begin{equation}
\left\{
\begin{array}{c}
\frac{\partial G}{\partial V} = \alpha G^{\prime}(x) \\
\frac{\partial G}{\partial T} = \beta G^{\prime}(x) 
\end{array}
\right.
\Rightarrow
\,
\frac{1}{G}\frac{\partial G}{\partial T} = k^{-1} \frac{1}{G} \frac{\partial G}{\partial V}
\end{equation}

\subsection{Benchmarking}
\begin{figure}[tpb]
\begin{centering}
\subfigure{\includegraphics[width=0.48\textwidth]{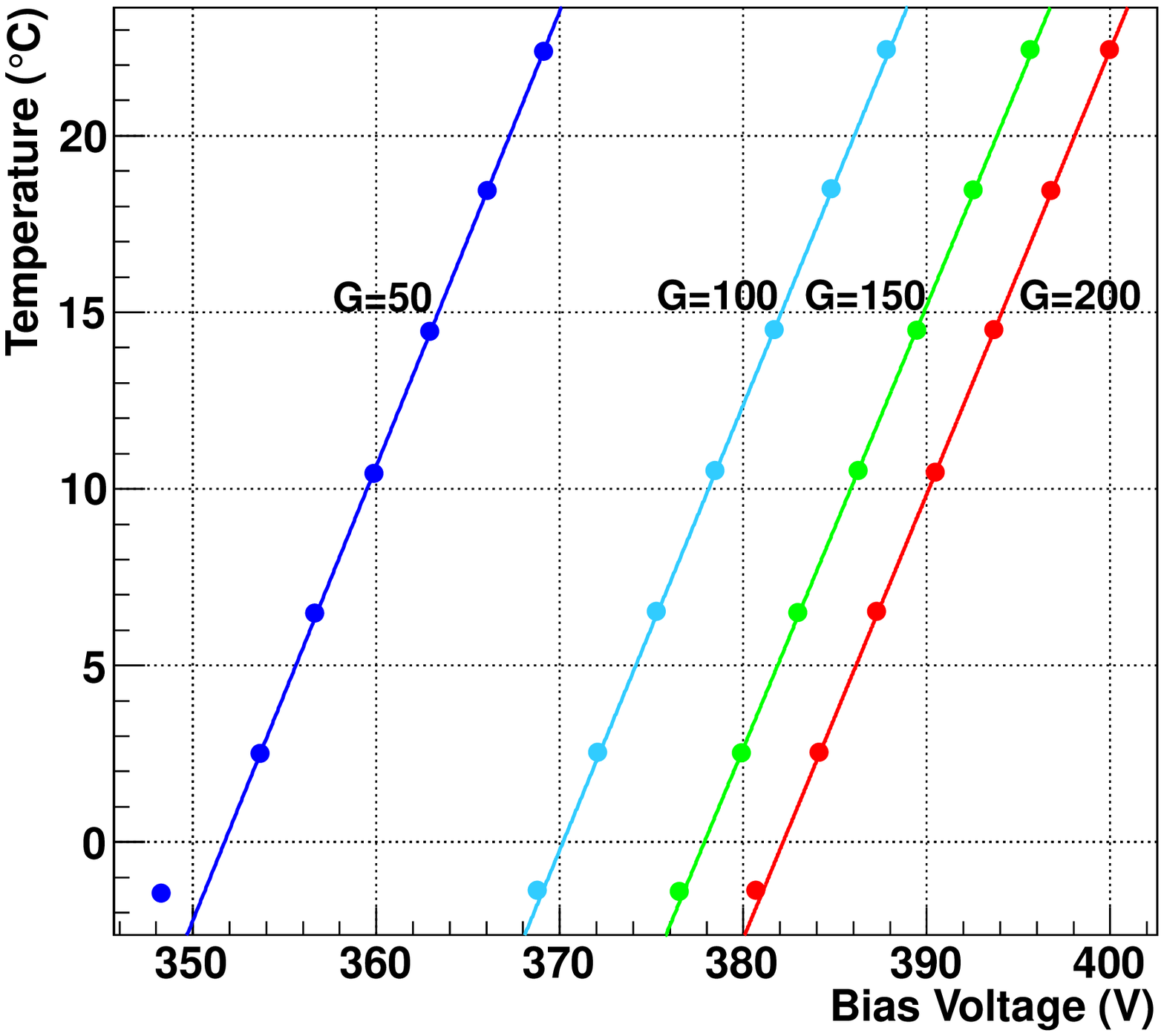}}\quad
\subfigure{\includegraphics[width=0.48\textwidth]{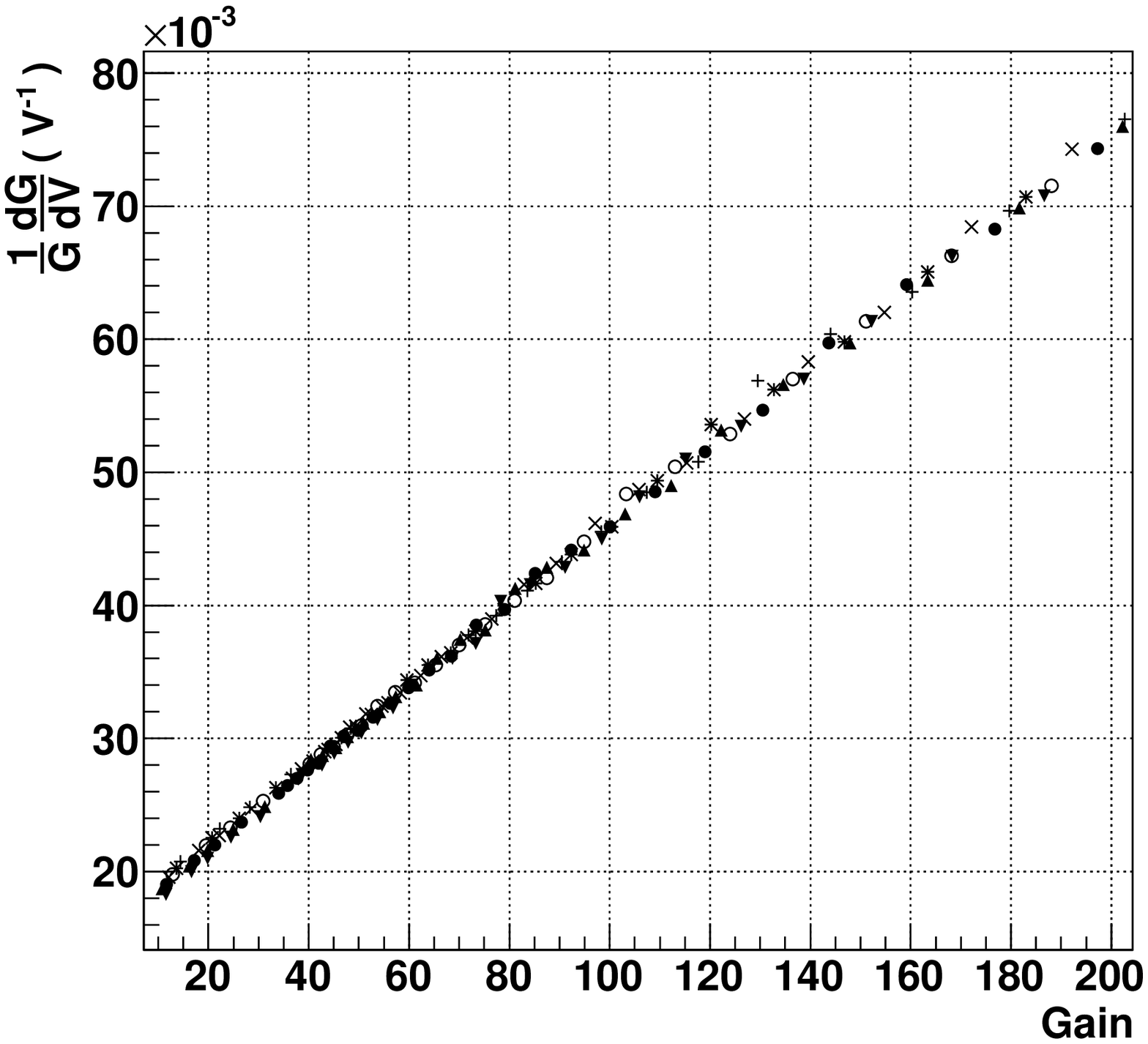}}
\caption{\label{fig:apdF6} \small Left: bias voltage corresponding to different LA-APD gain values, as a function of operating temperature. The curves are the results of the best fits performed with a linear function to the different data-sets to evaluate the LA-APD $k$ parameter (the lowest temperature point for $G=50$ was intentionally excluded from the fit). Right:  Measured $\frac{1}{G}\frac{dG}{dV}$ vs $G$ curves, for different operating temperatures. Same symbols as in the previous Figure.}
\end{centering}
\end{figure}

We identified a set of quality-checks to ultimately decide if a measured sensor is properly operational or not. Any APD not fullfilling any one of these requirements has to be characterized again and, if still presenting anomalies, discarded. 
\begin{itemize}
\item{The measured dark-current $I_{on}$ and the photo-current $I_{off}$ as a function of the bias voltage $V$ must be checked for all the measured temperatures. These data are required to have, qualitatively, the same behavior as reported in Figure~\ref{fig:apdF1}, i.e. the photo-current must show a plateau for $V \lesssim 50$ V, and then grow with approximately an exponential behavior. The dark-current $I_{off}$ should follow the $I_{on}$ behavior, and always be lower than it by at least one order of magnitude.}
\item{The dark-current behaviour $I_{off}$ as a function of the photo-detector gain $G$ has to be checked for all the measured temperatures. It should exibit a linear dependence, $I_{off} = c_1 + c_2 G$, and the constant contribution $c_1$ should decrease with the operating temperature.}
\item{The relative gain variation with respect to the bias voltage $\frac{1}{G}\frac{dG}{dV}$ must be finally checked, as a function of the gain $G$. Results obtained at different temperatures should approximately superimpose, for $G\gtrsim 10$, as follows directly from the gain functional dependence on bias voltage and temperature discussed before. A typical example is shown in Figure~\ref{fig:apdF6}, right panel.}
\end{itemize}

%\begin{figure}[tpb]
%\begin{centering}
%\includegraphics[width=0.48\textwidth]{fig10.eps}
%\caption{\label{fig:apdF6} Measured $\frac{1}{G}\frac{dG}{dV}$ vs $G$ curves, for different operating temperatures.
 %Symbols used are those described in Fig. \ref{fig:apdF8}.
%}
%\end{centering}
%\end{figure}

\section{Facility for APD characterization}\label{sec:Facility}

\begin{figure}[tpb]
\centering
\subfigure{\includegraphics[height=3.5in]{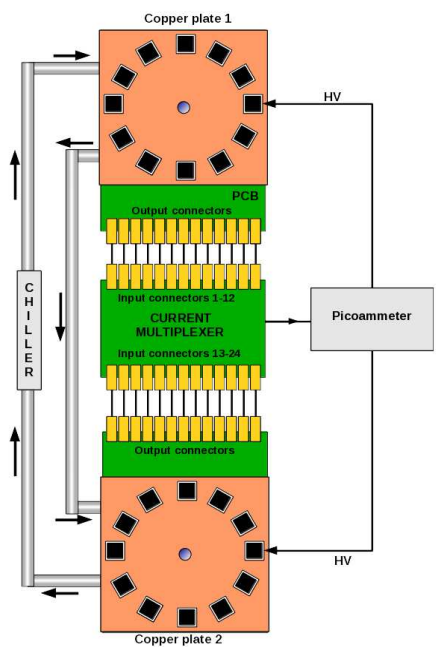}}\quad
\subfigure{\raisebox{15mm}{\includegraphics[width=0.48\textwidth]{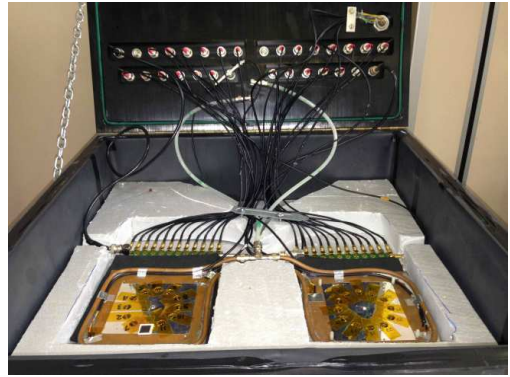}}}
\caption{\small Left: simplified scheme of the facility for LA-APD gain measurement. Right: picture of the facility, showing the two cold copper plates with holes for LA-APDs connection on the underlying PCB cards. One LA-APD is installed on the left copper plate.
%The part of the circuit to control the multiplexer is not shown.
}
\label{fig:Facility1}
\end{figure}

The layout of the facility for automatic APD characterization that we developed is shown in Figure~\ref{fig:Facility1}. The facility can measure up to 24 sensors at once, characterizing them in the voltage range between 0 and 500 V. The operating temperature can be varied from -2 $^{\circ}$C to +25 $^{\circ}$C. The measurement at each temperature takes approximately 2.5 hours. Therefore, one day is required to fully characterize 24 sensors from the minimum to the maximum temperature, with 8 measurements at $\simeq 3.5$ $^{\circ}$C intervals. The facility has been used to characterize the following APD models: Hamamatsu S8664-55 and Hamamatsu S8664-1010. There are no limitations to characterize other APD models, providing they have the same mechanical layout for the connection pins as the two aforementioned models.

\begin{figure}
\centering
\subfigure{\includegraphics[height=3in]{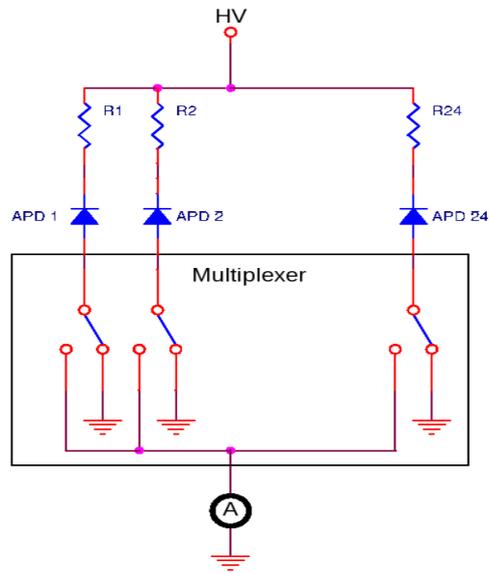}}\quad
\subfigure{\raisebox{5mm}{\includegraphics[height=2.5in]{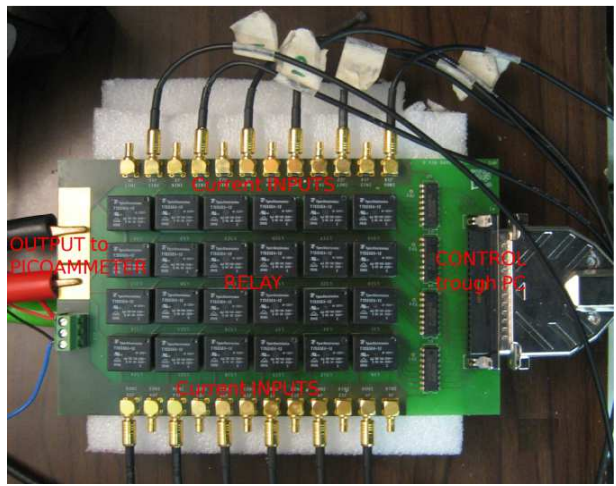}}}
\caption{\small Left: Schematic of the gain measurement circuit (just three channels are reported). ``HV'' is the common APDs high voltage source, while ``A'' is the picoammeter. Right: the current multiplexer.
}
\label{fig:apdF7}
\end{figure}

%\begin{figure}
%\begin{center}
%\includegraphics[width=0.48\textwidth]{apparatus.png}
%\caption{\small T}
%\label{fig:Facility2}
%\end{center}
%\end{figure}
Each installed sensor is characterized using the procedure described in the previous section. The dark current $I_{on}$ and the photo-current $I_{off}$ are measured for different bias voltages to obtain the internal gain. In order to employ a single picoammeter to measure the currents, and a single high-voltage source, a custom current multiplexer was developed. The current multiplexer is the core of the facility (see Figure~\ref{fig:apdF7}). During operation, the APD cathodes are all connected to the common HV source via independent bias resistors, while the anodes are connected to the multiplexer inputs. The multiplexer single output is connected to the picoammeter. In this way, only one APD at a time is physically connected to the instrument, while the other 23 are coupled to ground. We designed the current multiplexer using mechanical relays (type SPDT, ``single pole, double throw''\footnote{A SPDT relay is a 3-terminals device, with the common terminal connect to either of the two others, depending on the state of the magnetic coil.}), because the measured currents are of the order of 10-100~nA and other types of electronic switches, such as those based on transistors, would introduce spurious currents of the same order of magnitude, leading to incorrect gain measurements. Tyco ``T7SS5E4-12'' relays were selected because of the high isolation between the output contacts and the PCB mounting option. Relays are driven by four independent Darlington Transistor Arrays (Texas Instruments ULN2003A), that, in turn, are controlled trough a digital interface on the DAQ PC (NI 9403).

%\begin{figure}
%\begin{center}
%\includegraphics[height=2.4in]{fig8.eps}
%\caption{\small The current multiplexer of the facility for LA-APD Gain Measurement.}
%\label{fig:apdF8}
%\end{center}
%\end{figure}
The 24 APDs under measurement are mounted on two drilled copper plates, both hosting an LED in the center. Each sensor is mounted in correspondence to a hole, and connected to a PCB on the opposite side of the plate. The APDs are arranged circularly around the LED, which is also connected to the PCB. The PCB distributes the HV bias to each sensor, feeds their currents to the input of the current multiplexer, and provides current to the LED. The facility includes a cooling system to keep LA-APDs at a constant temperature. The coolant flows in a pipe connected to an external programmable chiller and welded to each of the two plates. The temperature is continuously monitored via 2 thermo-resistors mounted on each plate. The chiller provides temperature feedback loop, measuring the temperature of the return liquid, with 0.1 $^{\circ}$C stability. Furthermore, during operation, the copper assembly is inserted in a vacuum- and light-tight box flushed with nitrogen to prevent moisture formation.

The system, controlled via a Labview program running on the DAQ PC, works as follows. During the initial setup, the user enters the values for the temperature scan and the IDs of the sensors under measurement. For each programmed temperature, the APDs are individually enabled and characterized. At the end of the full cycle data are analyzed to calculate the gain of the APDs as a function of the bias voltage and temperature, the dark current dependence on the gain, and the relative gain variation at the different temperatures. The raw data and other relevant parameters, such as the bias voltage corresponding to the nominal gain at room temperature and the relative variation with respect to the temperature and the bias voltage in the neighborhood of the nominal working point, are then recorded to a file. A set of histograms is also produced, to let the user perform the quality check previously described.

\subsection{Systematic effects}
During the facility commissioning we investigated the relevance of possible systematic effects in the measurement, such as those due to long term fluctuations introduced by temperature drifts, residual moisture formation, and variation in the LED emission.
The gain of the same LA-APD has been repeatedly measured over a time period of 15 days, and the bias voltage corresponding to a fixed gain at room temperature ($G=150 \, @ \, T=+18$ $^{\circ}$C for the specific sensor) has then been calculated for each measurement. Results showed that the working point was constant within less than 0.1 V for all measurements.
%Results are reported in Fig. \ref{fig:apdF7}: all the measurements show the working point within less than 0.1 V.
This corresponds to a relative error on the measured gain approximately equal to $0.6 \% $, being $\frac{1}{G}\frac{\partial G} {\partial V} \simeq 6 \% / V $ in the neighborhood of the LA-APD selected working point ($G=150$). The same systematic effect was found on the other measured quantities, such as the dark current and the $k$ parameter.

\section{FT-Cal LA-APD characterization results}

\begin{figure}[tpb]
\centering
\includegraphics[width=0.47\textwidth]{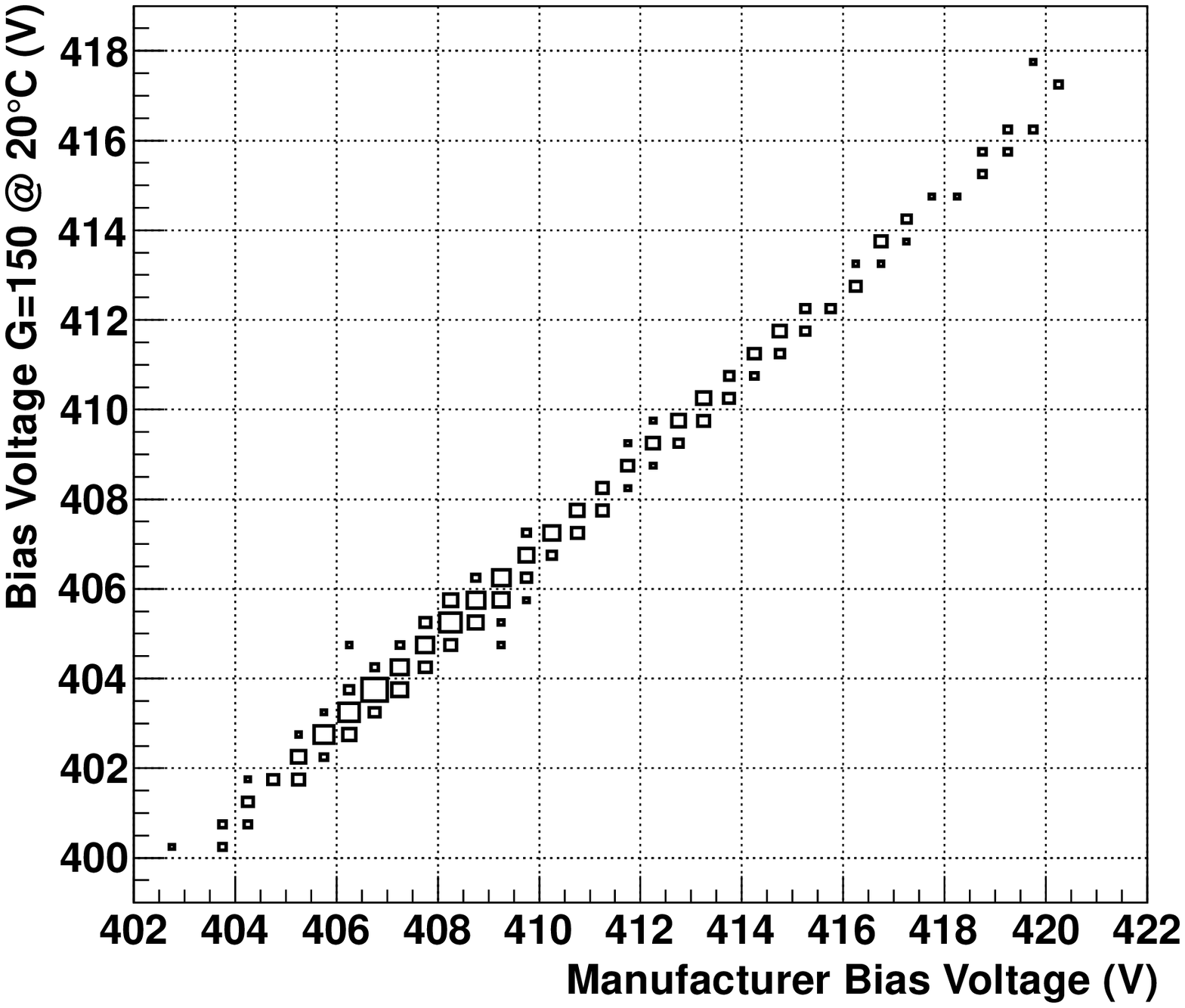}
\includegraphics[width=0.47\textwidth]{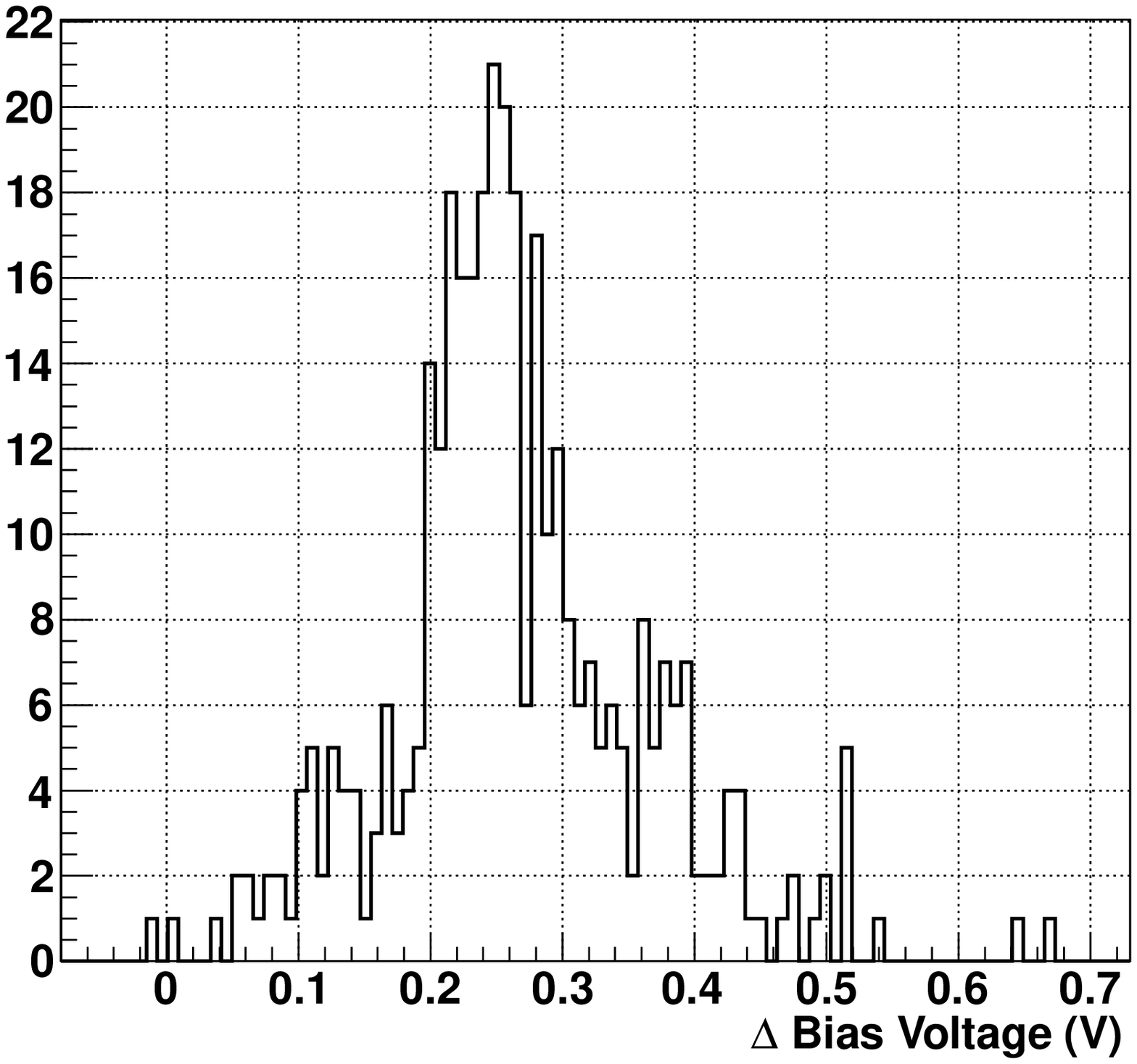}
\caption{\small \label{fig:APDF9}
Left: correlation between the \textit{measured} bias voltage for $G=150$ at $+20$ $^{\circ}$C ($y$ axis) and the value reported by the manufacturer for $G=50$ at $+25$ $^{\circ}$C ($x$ axis). Right: absolute bias voltage difference for $G=150$ between the measure performed at $0$ $^{\circ}$C and the extrapolated result from $+20$ $^{\circ}$C.}
\end{figure}

The first large-scale use of the APD characterization facility was to measure the LA-APD employed in the FT-Cal detector of the CLAS12 experiment at Jefferson Laboratory. The 380 sensors were characterized using the above facility within one month. Results were analyzed to identify and select the properly working sensors. As a result, 17 samples were discarded, since they did not fulfill the previously described data-quality checks, either after a second measurement. Figure \ref{fig:APDF9}, left panel, shows the comparison between the measured bias voltage corresponding to APD gain $G=150$ at $+20$ $^{\circ}$C, and the value declared by the manufacturer for the working point $G=50$ at $+25$ $^{\circ}$C, for the accepted samples.
The good correlation demonstrates the reliability of the system. 
The right panel, instead, shows the absolute difference between the bias voltage measured for $G=150$ at 0 $^{\circ}$C and the corresponding value extrapolated from the measure at  $20$ $^{\circ}$C, using the procedure previously described. The average difference of $\simeq$ 0.25 V corresponds approximately to a gain uncertainty of $\simeq 2\%$ (see Figure \ref{fig:apdF6}). This is a worst-case scenario, since in reality APDs are characterized at temperatures close to the foreseen working point, thus reducing the extrapolation error. The result, however, proves the reliability of the gain extrapolation procedure in the whole $(V,T)$ parameter space.

\section{Conclusions}

We developed a facility to automatically characterize Avalanche PhotoDiodes in the -2 $^{\circ}$C - 25 $^{\circ}$C temperature range. Such a facility can characterize up to 24 sensors at once, within one day of operation, measuring APDs from the minimum to the maximum temperature in $\simeq 3.5$ $^{\circ}$C intervals . For each sensor, the bias voltages corresponding to the nominal gain of 150 at different temperatures, and the relative gain variation with respect to the voltage and temperature are recorded. 
Particular care was taken to check possible systematics on the measured gain, that have been evaluated to be of the order of $0.6\%$, for a sensor gain $G=150$.

\section{Acknowledgments}

The authors want to acknowledge the electronic group of INFN-Genova section for the support given during the design of this facility and A. Balbi for help during the construction.

%%
%% End of file `elsarticle-template-1-num.tex'.

\begin{thebibliography}{9}

\bibitem{CMS-ECAL}
CMS Collaboration,
\textit{Performance and operation of the CMS electromagnetic calorimeter},
\emph{Journal of Instrumentation} \textbf{5} (2013) T03010

\bibitem{PHOS-ECAL}
D.C. Zhou (for the ALICE Collaboration), \textit{PHOS, the ALICE-PHOton spectrometer},
\emph{Journal Of Physics G: Nuclear and Particle Physics} \textbf{34} (2007) S719

\bibitem{PANDA-ECAL}
W. Novotny et al. (PANDA collaboration), \textit{The Electromagnetic Calorimetry of the PANDA Detector at FAIR}, \emph{Journal of Physics: Conference Series} \textbf{404} (2012) 012063


\bibitem{FT-CAL}
  A.~Celentano et al. (CLAS Collaboration),
  \textit{The Forward Tagger facility for low Q$^2$ experiments at Jefferson Laboratory}, \emph{EPJ Web Conf}  {\bf 73} (2014) 08004.

\bibitem{HPS-CAL} 
  M.~Battaglieri et al. (HPS Collaboration),
   \textit{The Heavy Photon Search Test Detector}, arXiv:1406.6115 [physics.ins-det].
 

\end{thebibliography}
\end{document}